\documentstyle[aps,epsfig,twocolumn,floats,array]{revtex}

\title{Dynamics of evaporative cooling in magnetically 
trapped atomic hydrogen}

\author{Makoto Yamashita,$^1$ Masato Koashi,$^2$ Tetsuya Mukai,$^1$ 
Masaharu Mitsunaga,$^{1,*}$ 
and Nobuyuki Imoto$^{1,2}$\\}
\address{
$^1$NTT Basic Research Laboratories, 3-1, Morinosato Wakamiya, 
\\ Atsugi-shi, Kanagawa 243-0198, Japan 
}
\address{
$^2$The Graduate University for Advanced Studies, Shonan Village, \\
Hayama, Kanagawa 240-0193, Japan
}
\date{\today}

\begin{document}
%\draft
\maketitle

\begin{abstract}
We study the evaporative cooling of magnetically trapped atomic 
hydrogen on the basis of the kinetic theory of a Bose  gas.   
The dynamics of trapped atoms is described by the coupled 
differential equations, considering both the evaporation and 
dipolar spin relaxation processes.  
The numerical time-evolution calculations quantitatively 
agree with the recent experiment of
Bose-Einstein condensation with atomic hydrogen. 
It is demonstrated that the balance between 
evaporative cooling and heating due to dipolar relaxation limits   
the number of condensates to $ 9  \times 10^8 $  and the corresponding 
condensate fraction to 
a small value of 4\% as observed experimentally. 
\end{abstract}

\bigskip

Pacs. 03.75.Fi, 05.20.Dd, 32.80.Pj
\renewcommand{\thefootnote}{\fnsymbol{footnote}}
\footnotetext[1] {Present address:  Department of Physics, Faculty of Science, Kumamoto University,  
Kurokami 2-39-1, Kumamoto 860-8555, Japan.}

\vspace{1cm}
The recent development of atom-manipulation techniques \cite{Manipu} 
has realized Bose-Einstein condensation
(BEC) in magnetically trapped alkali-metal atoms \cite{Rb,Li,Na} and 
atomic hydrogen \cite{H}.  In all these
experiments, evaporative cooling was adopted at the final stage of 
the cooling procedures and was essential in obtaining the extremely low 
temperatures needed for the quantum degeneracy. 
The cooling mechanism of this powerful method is based on both the 
selective removal of energetic atoms through evaporation and collisional
rethermalizations among the remaining atoms \cite{Evap}. 
Evaporative cooling itself is 
stimulating theoretical studies of the process 
of condensate formation in nonequilibrium atomic gases
 \cite{Luit,Berg,Sackett,Gar,Holl,Wu,Jaksch,1D_ev,Yama,Walser,Drum}.
\par 
In BEC experiments with atomic hydrogen, evaporative cooling has been 
implemented by lowering the potential height of the saddle point at 
one end of a magnetic trap with cylindrical symmetry \cite{Manipu}. 
This approach suffers from a reduction of the dimension of evaporation 
at low temperatures due to low elastic collision rate
\cite{Cilli_less,1D_ev} since the $s$-wave scattering length, $a$,  
of a hydrogen atom is anomalously small, 
about two orders of magnitude smaller than that in alkali-metal atoms.
Heating caused by the dipolar spin relaxation then easily retards the 
evaporative cooling and prevents further cooling before reaching the 
critical temperature of BEC. 
To overcome this problem, Fried {\it et al.} adopted the evaporative 
cooling induced by radio-frequency (rf)
magnetic field just after  the ^^ ^^ saddle-point" evaporative cooling \cite{H}. 
The atom ejection technique utilizing the transition between trapped and 
untrapped spin states enables efficient three-dimensional evaporation 
even at low  temperatures and in a highly anisotropic magnetic 
potential \cite{Evap}. 
BEC was finally achieved as a result of the heating-cooling balance in this 
efficient ^^ ^^ rf-induced" evaporative cooling. 
\par
There are several characteristic features of BEC in atomic hydrogen as 
compared with that in alkali-metal atoms \cite{H}. The small mass, $m$, 
of hydrogen resulted in the transition temperature of $50 \ \mu$K, 
the highest among BEC experiments. 
While the condensate fraction, i.e., the ratio of 
condensates to the whole trapped atoms, remained
a small value of about $5$\%, a huge number of 
condensates containing about $10^9$ atoms were observed and 
the peak density of the condensates was about 25 times higher 
than that of the noncondensed atoms.  
Furthermore, with respect to the loss of 
condensates, both the three-body recombination process and 
the background gas collisions in a cryogenic 
environment of trap are negligible in the hydrogen
system. Only dipolar spin relaxation becomes the dominant loss 
mechanism, and causes serious heating \cite{H_spec,dipolar}.
\par 
In this paper, the kinetic theory for evaporative cooling of a magnetically 
trapped Bose gas \cite{Yama} is extended to include dipolar relaxation loss. 
This theory enables the consistent and quantitative investigations on the whole evaporative 
cooling process of the experiments, from the classical regime at the early 
stage of cooling to the quantum degenerate regime after the BEC 
transition. 
We applied it to the rf-induced evaporative cooling 
in the recent BEC experiment with atomic hydrogen \cite{H} and 
quantitatively investigated the characteristic features of hydrogen BEC 
mentioned above. 
We expect further that our calculation studying the dynamics 
of trapped atoms in cooling process will give a good estimation 
for the growing of condensates in the experiment, since 
there is room for discussions on the number of condensates measured 
in the spectroscopic way \cite{H}. 
\par
Here, we briefly mention the formulation of the theory \cite{Yama}. 
During evaporative cooling, the applied rf-magnetic field effectively 
truncates the trapping potential, $U({\bf r})$, at the energy, $\epsilon_t$, 
determined by the rf-field frequency. 
The thermalized distribution of noncondensed atoms 
in such a truncated potential is well
approximated by the truncated Bose-Einstein distribution function 
\begin{equation}
\tilde f({\bf r}, p)=\frac{1}
{\tilde \xi ^{-1} ({\bf r}) \cdot \exp (\epsilon_p/k_BT) -1}\cdot \Theta 
(A({\bf r})-\epsilon_p), 
\label{BE_dis}
\end{equation} 
where $T$ is the temperature, $\epsilon_p=p^2/2m$ is the kinetic energy 
of atoms, and $\Theta (x) $ is the step
function. $\tilde\xi({\bf r})$ denotes the local fugacity including the 
mean field potential energy; 
$\tilde \xi({\bf r}) =\exp \{[\mu - U({\bf r})-2v\tilde n({\bf r})]/k_BT\}$, 
where $\mu $ is the chemical potential, $v=4\pi a \hbar^2/m$ is interaction 
strength between trapped atoms, and
$\tilde n({\bf r})$ is the density profile of atoms in a truncated potential. 
The step function in Eq.\ (\ref{BE_dis}) eliminates the momentum states 
whose kinetic energies exceed the 
effective potential height
$A({\bf r})=\epsilon_t-U({\bf r})-2v\tilde n({\bf r}) $. The momentum 
integration of this truncated
distribution function gives the density profile in a self-consistent way such 
that  $\tilde n({\bf r})=4\pi h^{-3}\int
\tilde f({\bf r}, p ) \ p^2 {\rm d} p$, and similarly the internal energy density  
such that 
$\tilde e ({\bf r})=4\pi h^{-3}\int \epsilon_p \tilde f({\bf r}, p ) \ p^2 {\rm d} p 
+v\tilde n^2({\bf r}) 
+U({\bf r}) \tilde n({\bf r})$. The total number of atoms, $\tilde N$, and the 
total internal energy, $\tilde E$, in a 
truncated potential are evaluated respectively by the spatial integrations of 
these density functions,  
$\tilde N=\int \tilde n({\bf r}) {\rm d}{\bf r}$ and 
$\tilde E=\int \tilde e({\bf r}) {\rm d}{\bf r}$. 
At low temperatures, the density profile in the condensed region is 
described as the sum of condensates, 
$n_0({\bf r})$, and saturated noncondensed atoms, $\tilde n_n$, 
such that $\tilde n({\bf r})=n_0({\bf r})+\tilde n_n$. 
Condensates obey the Thomas-Fermi distribution $n_0({\bf r})=
n_p-U({\bf r})/v$, where $n_p$ 
represents the peak density of condensates at the center of the 
potential. The density $\tilde n_n$ 
is evaluated through the truncated  Bose-Einstein distribution 
function under the condition $\tilde \xi({\bf r})=1$. 
\par 
The dynamics of trapped atoms during evaporative cooling is investigated 
on the basis of kinetic theory of a Bose gas \cite{Yama}.  
Hydrogen atoms are removed from the trapping potential through both 
evaporation and  dipolar spin relaxation 
processes \cite{H,H_spec,dipolar}.  The change rates (i.e., loss rates) 
of density functions, $\dot n_{\rm loss}$ 
and $\dot e_{\rm loss}$, are evaluated respectively as the sum of 
the contributions of both processes, such that 
\begin{eqnarray}
\nonumber \dot n_{\rm loss}({\bf r}) &=& 
- \dot n_{\rm ev }({\bf r})- G_2 \cdot K({\bf r}) \cdot 
\tilde n^2({\bf r}), \\
\dot e_{\rm loss}({\bf r}) &=& 
- \dot e_{\rm ev }({\bf r})- G_2 \cdot K({\bf r}) \cdot  
\tilde e({\bf r}) \cdot \tilde n({\bf r}),
\label{loss_den}
\end{eqnarray}
where $\dot n_{\rm ev}$ and $\dot e_{\rm ev}$ are the evaporation rates 
of density functions, $G_2$ is the dipolar 
decay rate constant, and $K$ is the correlation function which 
describes the second-order coherence of trapped atoms
 \cite{dipolar,kagan,stoof}.  Both $\dot n_{\rm ev}$ 
and $\dot e_{\rm ev}$ were 
derived from a general collision integral of a Bose gas system   
in Ref.\ \cite{Yama}, and we adopt here the opposite sign of the 
notations of these rates defined in this reference.  
The bosonic feature of evaporation process is therefore included 
as the strong enhancement of these evaporation rates 
due to the stimulated scattering of atoms. 
Regarding the dipolar spin relaxation 
process, we assume the correlation function of an ideal 
Bose gas such that 
$ K({\bf r})=(n_0^2({\bf r})+4n_0({\bf r}) \tilde n_n +2\tilde n_n^2)/2 
\tilde n^2({\bf r})$ \cite{dipolar,kagan,stoof}.  
The spatial integration of Eq.\ (\ref{loss_den}) gives important 
thermodynamic quantities such as the 
loss rate of total number of trapped atoms, $\dot N_{\rm loss}$, 
and that of the total internal energy,
$\dot E_{\rm loss}$:
\begin{eqnarray}
\nonumber \dot N_{\rm loss} &=& \int \dot n_{\rm loss} ({\bf r})
\ {\rm d \bf r}=-\dot N_{\rm ev} - \dot N_{\rm dip}, \\
\dot E_{\rm loss} &=&  \int \dot e_{\rm loss}  ({\bf r})\  {\rm d \bf r}=
- \dot E_{\rm ev} - \dot E_{\rm dip}, 
\label{loss_whole}
\end{eqnarray} 
where $\dot N_{\rm ev} (\dot E_{\rm ev}) $ and  $\dot N_{\rm dip} 
(\dot E_{\rm dip}) $ represent the evaporation rate and
dipolar relaxation loss rate, respectively. 
These loss rates are complicated functions of temperature $T$,  chemical 
potential $\mu$, and truncation energy $\epsilon_t$. 
Finally, the system obeys the coupled differential equations 
\begin{equation}
\frac{\partial \tilde N}{\partial t}=\dot N_{\rm loss}, \hspace{10mm} 
\frac{\partial \tilde E}{\partial t}=
\dot E_{\rm loss}.
\label{kine_eq}
\end{equation}
Time-evolution calculations were performed numerically on the 
assumption of quick rethermalization, which is appropriate for 
the slow evaporative cooling used in the 
usual BEC experiments \cite{Yama}. Thus, the system always stays 
in a quasi-thermal equilibrium state and is 
described by the truncated Bose-Einstein distribution function 
in Eq.\ (\ref{BE_dis}). 
\par   
Precise knowledge of experimental conditions is also necessary 
for our quantitative calculations. 
The magnetic trap adopted in the BEC experiment with atomic hydrogen
\cite{H} is  modeled by the Ioffe-Pritchard potential
$U(\rho,z)=\sqrt{(\alpha \rho)^2+(\beta z^2+\theta)^2}-\theta$ 
with the radial potential gradient $\alpha=2.2\times 10^{-23}$ J/m, 
the axial curvature $2\beta=6.8\times 10^{-24} \ {\rm J/m}^2$,
and the bias energy $\theta=4.8 \times 10^{-28}$ J (i.e., 
$\theta/k_{\rm B}=35\ \mu $K).  
At low energies, this potential is well
approximated by the harmonic one with the radial oscillation frequencies
$\omega_\rho=\alpha/\sqrt{m\theta}=2\pi \times 3.9 $ kHz and the axial 
frequency $\omega_z=\sqrt{2\beta/m}=2\pi \times 10$ Hz. 
It should be noted here that, even after the BEC transition, many 
noncondensed hydrogen atoms still distribute in the 
high-energy region, where such harmonic approximation fails.  
We treated the Ioffe-Pritchard potential exactly for the quantitative 
description of noncondensed atoms.  Accordingly, all density functions 
were calculated in the cylindrical coordinate. 
\par
As is well known, the efficiency of evaporative cooling strongly depends on 
the way the rf-field frequency is swept 
(i.e., forced evaporative cooling) \cite{Evap}. 
The high efficiency is obtained by the slow evaporative cooling which 
continues for the time on the order of 10 seconds.  In the BEC experiment 
with atomic hydrogen, the rf-field
frequency was swept according to the functional form \cite{willmann}  
$\displaystyle \nu(t)=\nu_{\rm i} \left(
\frac{\nu_{\rm f}}{\nu_{\rm i}}\right)^{\tau^g} \label{freq_time}$, 
where $\nu_{\rm i}=23$ MHz is the initial rf-frequency, 
$\nu_{\rm f}=2$ MHz is the final one,  
$\tau=t/t_0$ denotes the normalized time with the specific 
time $t_0=25$ s, and $g=1.5$ is the fixed parameter. 
The truncation energy $\epsilon_t$ then changes according to the 
relation $\epsilon_t(t)=h\nu(t)-\theta$. 
\par
Furthermore, we employed the following experimental parameters \cite{H}: 
{\it s}-wave scattering length $a=6.48 \times 10^{-2}$ nm \cite{H_scatt},  
the dipolar 
decay rate constant $G_2=1.1\times10^{-15}$ cm$^3$/s \cite{H_spec}, 
initial temperature $T=120 \ \mu$K,   
initial peak density $\tilde n ({\bf 0})=5.0\times 10^{13}$ cm$^{-3}$,
and the initial number of trapped atoms in a truncated 
potential $\tilde N =1.16 \times 10^{11}$. 
The time-evolution of evaporative cooling was calculated for 25 s, 
i.e., the normalized time $\tau$ changes from 0 to 1. 
\begin{figure}[htb]
  \begin{center}
    \leavevmode
 \begin{minipage}{6.8cm}
    \epsfig{file=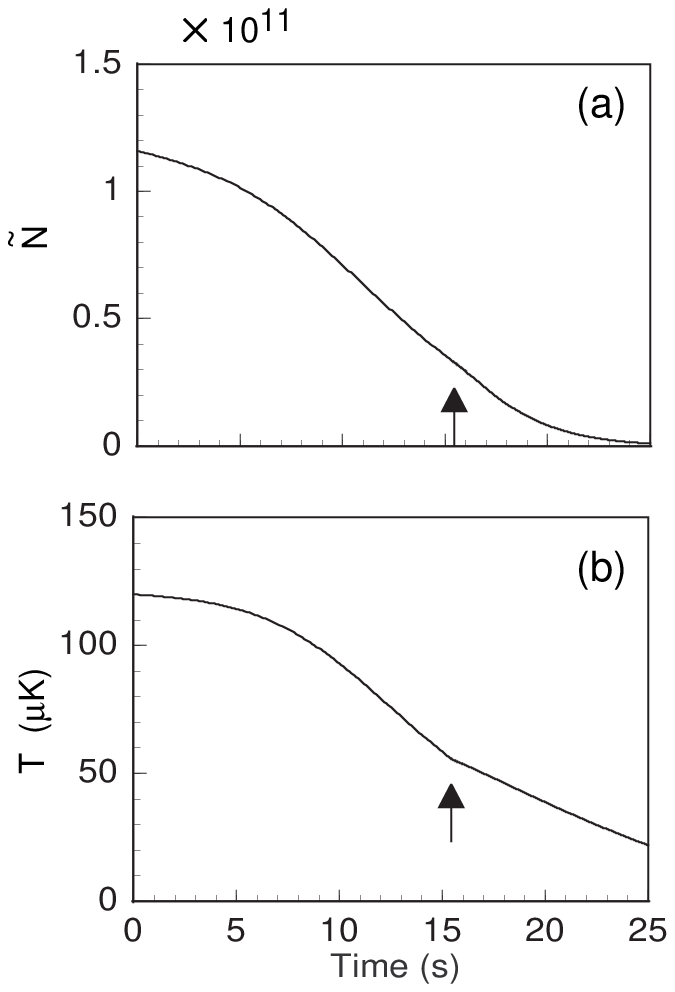,angle=0,width=\linewidth}
 \end{minipage}
  \end{center}
  \caption{Time evolution of (a) the total number of trapped atoms $\tilde N$, 
and (b) temperature $T$. 
The arrow indicates the point at which the BEC transition occurs. 
The heating due to the dipolar spin relaxation occurs after the 
BEC transition.
}
  \label{carrierfig}
\end{figure}
\par
In Figs.\ 1(a) and 1(b), we show the time evolution 
of the total number of trapped atoms $\tilde N$ and that of 
temperature $T$, respectively. 
$\tilde N$ decreases monotonically from $1.16\times 10^{11}$ 
to $8.7\times10^{8}$ during a 25-s evaporative cooling. 
The BEC transition occurs at 15.4 s ($\tau=0.62$) as indicated by the arrow 
in the figure.  
The decrease of temperature (from 120 to $22 \ \mu$K) is not so large in 
comparison with that has been demonstrated in alkali-metal 
atoms \cite{Rb,Li,Na,Evap}. 
We can see that the cooling speed strongly slows down after the BEC 
transition in Fig.\ 1(b), indicating considerable heating occurs.   
Here we show the calculated results at the BEC transition point:
temperature $T=56 \ \mu$K, peak density $\tilde n ({\bf 0})=2.0\times 
10^{14}$ cm$^{-3}$, number of trapped atoms  $\tilde N =3.3\times 10^{10}$, 
and truncation energy $\epsilon_t/k_{\rm B}=302\ \mu$K. 
These parameters are quite consistent with the experiment \cite{H}, 
which confirms the validity of our numerical calculations. 
\par
\begin{figure}[htb]
    \leavevmode
  \begin{center}
 \begin{minipage}{6.0cm}
    \epsfig{file=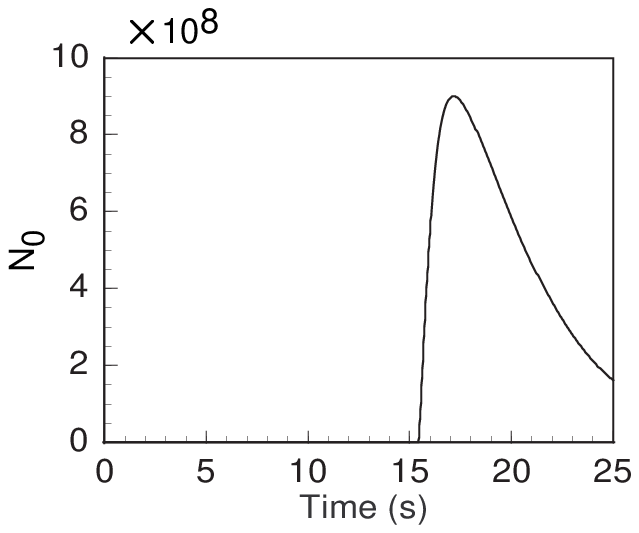,angle=0,width=\linewidth}
 \end{minipage}
  \end{center}
  \caption{Time evolution of (a) the total number of trapped atoms $\tilde N$, 
and (b) temperature $T$. 
The arrow indicates the point at which the BEC transition occurs. 
The heating due to the dipolar spin relaxation occurs after the 
BEC transition.}
\end{figure}

The condensate formation process shows the characteristic feature 
depicted in Fig.\ 2.  
Condensates first grow very rapidly after the transition. 
We can explain this by the fact that bosonic stimulation in evaporative 
cooling process \cite{Yama} strongly
accelerates the growing speed,  since the density of condensates 
becomes very high in atomic hydrogen.  
The number of condensates $N_0$ reaches the maximum value at 
17.2 s ($\tau=0.69$), and then decreases gradually due to dipolar decay. 
The calculated parameters at this maximum point are:
temperature  $T=49 \ \mu$K, peak density of condensates $ n_p=
4.3\times 10^{15}$ cm$^{-3}$, 
peak density of noncondensed atoms $\tilde n_n =1.7
\times 10^{14}$ cm$^{-3}$, 
total number of condensates $N_0=9.0 \times 10^{8}$, 
total number of noncondensed atoms $\tilde N_n=2.0 \times 10^{10}$, 
and truncation energy $\epsilon_t/k_{\rm B}=239\ \mu$K.  
The corresponding condensate fraction at the maximum point of $N_0$ 
is calculated as $f=N_0/(N_0+\tilde N_n)$ to be a small value of 4.3\%. 
These results quantitatively agree with the recent observation of BEC 
with atomic hydrogen \cite{H}. 
\par
The number of trapped atoms has been evaluated experimentally by 
measuring the density-dependent frequency shift of the two-photon 
1{\it S}-2{\it S} transition in atomic hydrogen \cite{H_spec}. 
The condensate number, $N_0\simeq10^9$ in Ref. \cite{H},  
was obtained on the assumption that the frequency shift for condensates 
is as large as that for noncondensed atoms with the same density. 
If we expect the disappearance of exchange effects in the 
excitation of condensates as discussed in Ref.\cite{H}, 
the frequency shift for condensates becomes half of that for 
noncondensed atoms, and many more condensates, $N_0\simeq 6\times 10^9$, 
with an unreasonably higher condensate fraction, $f=25\%$, are given. 
The calculated results in Fig.\ 2  support the evaluation of condensate 
number which assumes the equal frequency shift for condensates and  
noncondensed atoms. 
\begin{figure}[htb]
  \begin{center}
    \leavevmode
 \begin{minipage}{7.0cm}
    \epsfig{file=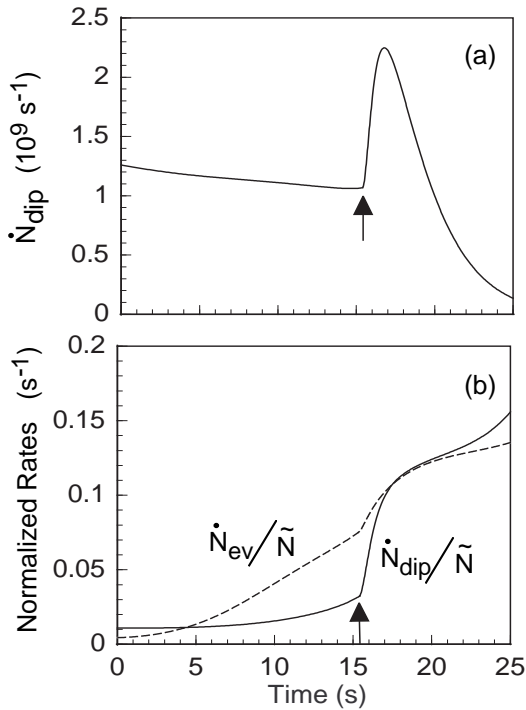,angle=0,width=\linewidth}
 \end{minipage}
  \end{center}
  \caption{Time evolution of (a) the dipolar loss rate  
$\dot N_{\rm dip}$, and (b) normalized rates, 
$\dot N_{\rm ev}/\tilde N$ (dashed line) 
and $\dot N_{\rm dip}/\tilde N$ (solid line).  
The dipolar loss rate exceeds the evaporation rate 
at around 17.5 s.
The arrow indicates the point at which the BEC 
transition occurs.
}
  \label{lossrate}
\end{figure}
\par
Next, we discuss the influence of dipolar loss on evaporative cooling.  
From Eq.\ (\ref{loss_den}), the dipolar relaxation process preferably 
removes atoms in the region with higher density. 
Since the removed atoms have lower internal energy in the trapping 
potential, the dipolar relaxation loss results in system heating.  
The time evolution of the dipolar loss rate 
$\dot N_{\rm dip}$ is shown in Fig.\ 3(a). 
The curve exhibits the jump structure around 
the BEC transition point indicated by the arrow.
  The serious heating after the transition in Fig\ 1(b) is caused by 
such drastic increase of dipolar loss known 
as ^^ ^^ relaxation explosion" in hydrogen system \cite{dipolar}. 
To clearly demonstrate the competition between evaporation and 
dipolar relaxation processes during evaporative 
cooling, the normalized rates $\dot N_{\rm ev}/\tilde N$ and  
$\dot N_{\rm dip}/\tilde N$ are plotted 
together in Fig.\ 3(b). 
The evaporation rate is larger than the dipolar loss rate until  
just after the BEC transition, which indicates that evaporative 
cooling is sufficiently efficient in this region. 
The explosively enhanced dipolar-loss rate after the transition 
finally exceeds the evaporation rate at around 17.5 s, 
and the resultant heating becomes serious. 
We note here that this time almost corresponds to the point 
where the number of condensates shows the maximum 
value in Fig.\ 2. The heating-cooling balance therefore limits 
the producible number of condensates and the
corresponding condensate fraction. 
One can expect that the trapping potential with weaker confinement 
and the optimized evaporative cooling will move this balance 
in the direction of a larger condensate production \cite{willmann}. 
\par
In conclusion, we have investigated the rf-induced evaporative 
cooling of magnetically trapped atomic 
hydrogen on the basis of the kinetic theory of a Bose gas. 
This approach can be applied over the whole temperature region of 
evaporative cooling while the earlier investigations based on the 
classical kinetic theory \cite{Luit,Berg,Sackett,Cilli_less,dipolar} 
fail around the critical point of BEC transition.
The calculated results quantitatively agree with 
the recent BEC experiment of atomic hydrogen.  This proves our 
calculations very useful, and the important 
future work would be to optimize the cooling trajectory 
for a larger production of condensates \cite{Sackett}. 
Our theory, on the other hand, assumes that the system 
always stays in
the quasi-thermal equilibrium states during evaporative cooling. 
A precise study of the deviation from the quasi-thermal 
equilibrium \cite{H_droplet} would give the limit 
of the approximation in the theory. 
\par
We thank L. Willmann of MIT for his valuable 
suggestions and discussions.

\newcounter{q}
\setcounter{q}{118}

\end{document}